\documentstyle[12pt,epsf]{article}
\input epsf

\newcommand {\cen}[1]{\begin{center} #1 \end{center}}
\newcommand {\eq} {\begin{equation}}
\newcommand {\qe} {\end{equation}}
\newcommand {\eqa} {\begin{eqnarray*}}
\newcommand {\qea} {\end{eqnarray*}}
\newcommand {\eqan} {\begin{eqnarray}}
\newcommand {\qean} {\end{eqnarray}}
\newcommand {\eqn} {\begin{displaymath}}
\newcommand {\qen} {\end{displaymath}}
\newcommand {\lp} {\left(}
\newcommand {\rp} {\right)}

\newcommand {\lc} {\left\{}
\newcommand {\rc} {\right\}}

\newcommand {\lcurly} {\mbox {\Large\{ }}
\newcommand {\rcurly} {\mbox {\Large\} }}

\newcommand {\lbrac} {\mbox {\Large[ }}
\newcommand {\rbrac} {\mbox {\Large] }}
\newcommand {\less} {{_{_{<}}}}
\newcommand {\great} {{_{_{>}}}}
\newcommand {\p} {^{\prime}}
\newcommand {\bigeta} {\mbox {\Large{$\eta$}}}
\newcommand {\bfq} {{\mathbf{q}}}
\newcommand {\bfqone} {{\mathbf{q_1}}}
\newcommand {\bfqtwo} {{\mathbf{q_2}}}
\newcommand {\bfqhat} {{\mathbf{\hat{q}}}}
\newcommand {\bfk} {{\mathbf{k}}}
\newcommand {\bfkone} {{\mathbf{k_1}}}
\newcommand {\bfktwo} {{\mathbf{k_2}}}
\newcommand {\bfkprime} {{\mathbf{k\p}}}

\newcommand {\bfkprimehat} {{\mathbf{ {\hat{k}} \p } }}
\newcommand {\bfkhat} {{\mathbf{\hat{k}} }}
\newcommand {\bfr} {{\mathbf r}}
\newcommand {\bfrone} {{\mathbf{r_1}}}
\newcommand {\bfrtwo} {{\mathbf{r_2}}}
\newcommand {\bfrhat} {{\mathbf{\hat{r}}}}
\newcommand {\bfronehat} {{\mathbf{ {\hat{r}}_1 } }}
\newcommand {\bfrtwohat} {{\mathbf{ {\hat{r}}_2 } }}
\newcommand {\bfrihat} {{\mathbf{ {\hat{r}}_i } }}
\newcommand {\bfkap}{\mbox {\boldmath $\kappa$}}

\newcommand {\bfsigmai} {\mbox {\boldmath $\sigma_i$}}
\newcommand {\bfsigmaone} {\mbox {\boldmath $\sigma_1$}}
\newcommand {\bfsigmatwo} {\mbox {\boldmath $\sigma_2$}}
\newcommand {\bfdel} {\mbox {\boldmath $\nabla$}}

\newcommand {\bfdelj} {\mbox {\boldmath $\nabla_j$}}
\newcommand {\bfdelone} {\mbox {\boldmath $\nabla_1$}}
\newcommand {\bfdeltwo} {\mbox {\boldmath $\nabla_2$}}
\newcommand {\Psil} {\Psi_l(kr)}
\newcommand {\Psip} {\Psi^{(+)}(\bfk,\bfrone)}
\newcommand {\Psim} {\Psi^{(-)\ast}(\bfkprime,\bfrtwo)}
\newcommand {\psione} {\psi(\bfk,\bfrone)}
\newcommand {\psitwo} {\psi(\bfkprime,\bfrtwo)}
\newcommand {\rhoone} {\rho(r_1)}
\newcommand {\rhotwo} {\rho(r_2)}
\newcommand {\rhoonesq} {\rho^2(r_1)}
\newcommand {\rhotwosq} {\rho^2(r_2)}
\newcommand {\rhopone} {\rho\p(r_1)}
\newcommand {\rhoptwo} {\rho\p(r_2)}

\newcommand {\fdsf}{F(\bfk,\bfkprime)}
\newcommand {\fdsfs}{F_S(\bfk,\bfkprime)}
\newcommand {\fdsfd}{F_D(\bfk,\bfkprime)}

\newcommand {\fdsfone}{F_1(\bfk,\bfkprime)}
\newcommand {\fdsftwo}{F_2(\bfk,\bfkprime)}
\newcommand {\pip}{$\pi^+$}

\newcommand {\pipm}{$\pi^{\pm}$}
\newcommand {\heth}{$^3$He}
\newcommand {\hth}{$^3$H}
\newcommand {\piphe}{$\pi^+$-$^3$He}
\newcommand {\pimhe}{$\pi^-$-$^3$He}
\newcommand {\pipmhe}{$\pi^{\pm}$-$^3$He}
\newcommand {\piph}{$\pi^+$-$^3$H}
\newcommand {\pimh}{$\pi^-$-$^3$H}
\newcommand {\pipmh}{$\pi^{\pm}$-$^3$H}

\newcommand {\pimn}{$\pi^-$-$n$}

\newcommand {\pipp}{$\pi^+$-$p$}
\newcommand {\degree}{$^{\circ}$}

\newcommand {\ea} {{\it et al.}}

\begin{document}

\cen{\large \bf Elastic Scattering of Pions From the 
Three-nucleon System}

\vspace*{.5in}

\cen{\large \bf S. L. Collier and W. R. Gibbs}

\cen{New Mexico State University} 
\cen{October 14, 1998}

\begin{abstract}

We examine the scattering of charged pions from the trinucleon system 
at a pion energy of 180 MeV.  The motivation for this study
is the structure seen in the experimental 
angular distribution of back-angle 
scattering for {\piphe} and {\pimh} but for neither {\pimhe} nor 
{\piph}. We consider the addition of a double spin flip term to 
an optical model treatment and find that, though the contribution 
of this term is non-negligible at large angles for {\piphe} and 
{\pimh}, it does not reproduce the structure seen in the experiment.

\end{abstract}

\newpage

\section{Introduction}
\begin{figure}
\epsfysize=6.0in 
\epsffile{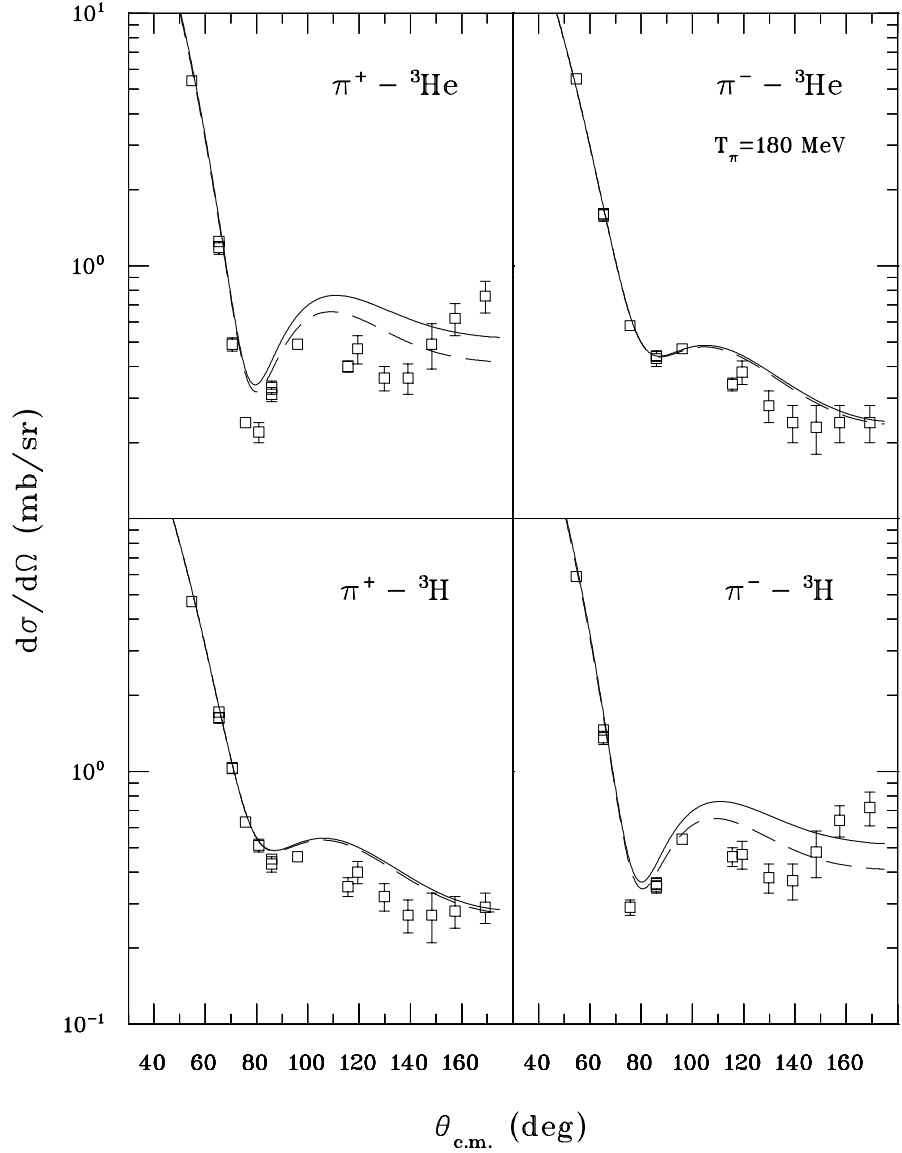}
\caption{Comparison of  experimental data from Ref.
\protect\cite{scott,data}
(squares) to a prediction from
the optical model of Ref. \protect\cite{gg} with DSF term (solid
line) and without DSF term (dashed line).}
\label{gr:ddsf}\end{figure}

The measurement of the differential cross section for the elastic 
scattering of {\pipm} from {\hth} and {\heth} has recently been
 extended into the backward
hemisphere by Matthews \ea, {\cite{scott}}. For all four cases, at an
incident pion energy of 180 MeV, 
there is a slight dip in the cross section at $\sim$~130{\degree} lab 
(or 5.8 fm$^{-2}$ momentum transfer), but for {\pimh}
and {\piphe} there is also a very distinctive rise in the cross
section at larger angles. Moreover, the authors stated 
that, due to the 
resolution of the spectrometer, the dip may be even deeper and 
sharper. They compared experimental data to scattering models by 
Kamalov, Tiator, and Bennhold {\cite{ktb}} and by Gibbs and Gibson 
{\cite{gg}}. For {\pimh} and {\piphe} both models agree reasonably 
well with the cross section up to 130{\degree}, but
both fail to predict the distinctive rise in the cross section at
large angles. For {\piph} and {\pimhe} both models give acceptable
agreement at all angles.  Figure~\ref{gr:ddsf} compares the experimental 
data \cite{scott,data} (squares) to a theoretical curve from Ref.
\cite{gg} (dashed line).

One is thus led to consider mechanisms that would give a significant 
rise in the cross section in the backward hemisphere for both 
{\pimh} and {\piphe}, but not for {\piph} or {\pimhe}.  Similar 
rises in large-angle cross sections have been seen for
{\pip} scattering from $^{12}$C and $^{16}$O 
\cite{burleson}. In the three-nucleon case the dependence of
this effect on the charge of the pion and on the target may
give an indication of the cause.

The pion kinetic energy, $T_{\pi}$=180~MeV, is near the peak of 
the $P_{33}$ resonance for $\pi$-nucleon scattering. At this 
energy the elastic scattering cross sections are in the ratio 
$\sigma (\pi^+p)/ \sigma(\pi^+n)
\approx \sigma (\pi^-n)/ \sigma (\pi^-p) \approx 9$. Therefore, 
we expect $\pi$-nucleus scattering at $T_{\pi}$=180~MeV to 
be dominated by the {\pipp} and {\pimn} amplitudes.
We use the convention 
that even-nucleon reactions refer to {\piphe} and {\pimh} reactions
(which have two dominant scattering centers), 
and odd-nucleon reactions refer to {\pimhe} and {\piph} reactions
(which only have one dominant scattering center at this energy).
We also use the nomenclature of Schiff \cite{schiff}
where an ``even'' or an ``odd'' nucleon corresponds, respectively, 
to one of the like nucleons or to the unlike nucleon (the neutron 
in {\heth} or the proton in {\hth}).

Qualitatively we can understand the first dip (around 90$^{\circ}$) 
as a minimum in the spin-independent amplitude due to the p-wave
dominance of the pion-nucleon phase shifts. 
For this intuitive view, consider the single-scattering impulse
approximation, where the basic dependence of the amplitude is given
by the sum of the relevant $\pi$N amplitudes multiplied by a
form factor. For $\pi$-nucleon scattering, the amplitude consists 
of two incoherent terms:  a non-spin flip term, $f(\theta)$, 
and a spin-flip term, $g(\theta)$.  Near the peak of the $P_{33}$ 
resonance $f(\theta) \sim \cos (\theta)$ (plus an s-wave contribution
which shifts the minimum slightly away from 90$^{\circ}$), and 
$g(\theta) \sim \sin (\theta)$.

In the approximation that the trinucleon system exists in a pure
s-state, (which we use throughout this paper), the amplitude $g(\theta)$
arises entirely from the interaction of the pion with the odd nucleon
since, in first order, the spin-flip amplitudes cancel from the
even nucleons. 
For \pimhe\  this interaction is strong, thus near 90\degree\ 
c.m. the minima in $|f(\theta)|^2$ is significantly filled in by 
$|g(\theta)|^2$. For the even-nucleon interaction {\piphe}, 
{\pipp} has the largest amplitude and the interaction with the
odd particle is weak so that the filling of the minimum is
considerably less. An analogous argument follows for {\pipmh}.  

It has been suggested {\cite{scotttogibbs}} that the rise in the
cross section at large angles might be due to the interaction of 
the incident pion with both of the like nucleons, 
flipping the spin of each so that the spin of the pair is conserved. 
That is, for {\pipmhe}, the incident pion sequentially induces a 
single spin flip of each proton, thereby leaving the final pair 
with spin 0.  
In the common optical model treatment, a potential is constructed from
the single scattering 
amplitude of the type discussed previously, and then the nuclear amplitude 
is obtained from the solution with this potential in a 
wave equation. Thus a spin projection change is absent from this type 
of treatment and must be calculated separately. For this 
reason double spin flip (DSF) 
scattering is not directly included in current optical models. The DSF, 
whose amplitude is coherent with $f(\theta)$, should clearly
have more effect on the cross section of the even-nucleon reactions than 
of the odd-nucleon reactions.  As each spin-flip amplitude preferentially 
leads to scattering around 90\degree, the two scatterings will 
lead to a forward-backward peaked angular distribution of the scattered
particles. The forward part will likely be much smaller 
than the amplitude from the first order optical potential, but at 
large angles the two contributions may be comparable.  

Franco \cite{franco} has investigated multiple spin flip
effects for $\pi$-$^4$He scattering in the Glauber approximation.
However, the Glauber approximation is not applicable in the large angle
scattering region where the DSF is expected to be of significance.

We treat the DSF as a second order correction to the
scattering model of Ref. \cite{gg}  and calculate it in the distorted wave
impulse approximation (DWIA).  This calculation is presented in
Section~\ref{sec:calc}, after briefly reviewing the scattering 
model of Ref.~\cite{gg} in Section~\ref{sec:om}. The results are 
discussed in Section~\ref{sec:results}, and conclusions are presented in
Section~\ref{sec:concl}.

\section{Basic Scattering Model}\label{sec:om}

In the case of moderately heavy nuclei the optical model can be
extended to include, by direct calculation, a number of effects
including finite range and medium modification of the pion-nucleon 
scattering amplitudes \cite{kg}.  The medium corrections can be 
understood, in an approximate sense, in terms of a shift in the
energy at which the phase shifts are to be evaluated and a 
transformation
from the pion-nucleon to the pion-nucleus frame (the so called ``angle
transform''). 
The effect of the finite range of the pion-nucleon
interaction is included by assuming plausible forms for the off-shell
t-matrix for the $\pi$N interaction.  A number of studies of the 
pion-nucleon interaction have attempted to determine the ranges (there is
a different range for each spin-isospin partial wave). For a recent
analysis using local potentials see Ref. \cite{gak}.  These ranges
can also be treated as phenomenological and fitted to the 
data\cite{elastic}.  

Another important element in pion scattering theory is due
the possibility of true absorption of the pion, converting
its mass into energy, as opposed to the usual ``optical model
absorption'' where the incident particle is simply removed from
the beam by inelastic scattering.  A number of attempts have been
made to include this effect from fundamentals \cite{garg,oset}. 
In the present calculations we adopt the method used in Ref. 
\cite{gg} of including an imaginary term in the potential 
proportional to the square of the density. 
Approximate values
of the parameter can be estimated from fits to heavier nuclei
(see Refs. \cite{elastic,dcx}).

The optical model contains a long known \cite{kmt} correction
due to the fact that, since the t-matrix is used to describe
each individual scattering, no nucleon can be struck twice successively.
Thus, in a multiple scattering picture, there can be A scatterings
the first time but only A--1 for each subsequent order.
If all scatterings are of the same strength,
this correction can be made by solving a wave equation 
with a potential having an overall strength of A--1, instead of A,  
and then multiplying the resulting 
t-matrix by the factor A/(A--1), (the KMT factor).
For an optical model in which there are only three nucleons involved,
this effect is of considerably greater importance than for a
heavier nucleus where this factor is close to unity.  We do not
attempt to do anything beyond what was used in Ref. \cite{gg}. 
The forward cross section
is normally the most reliably calculated in scattering theories and
it was noted in Ref. \cite{gg} that a treatment of this correction
involving factors of the order described above was necessary to obtain 
agreement with the data at forward angles. 

We have carried out a search over variations of the scattering
parameters used in Table~I of Ref.~\cite{gg}.  No substantial
rise in the backward direction was seen.  In order to focus on
the DSF contribution to the cross section, we have fixed the
values at:  Energy shift=0 MeV, angle transform parameter=1,
$\rho^2$ coefficient=8.8 fm$^4$, and the off shell ranges
(s and p) at 600 MeV/c.  For a discussion of the corrections which come
in for scattering of pions from few-nucleons systems 
see Ref. \cite{gghsk}.

For the KMT factor, the calculated value (Eq.~A4 of Ref.~\cite{gg})
is used. We emphasize
that a factor of this size is necessary in order to get agreement
in the forward direction.  If a factor of unity is used the cross
section forward of 90\degree\ is shifted toward smaller angles.  
However, see the discussion in the conclusion on this point.

The spin-flip amplitude, $g(\theta)$, is calculated in the
distorted wave Born approximation, where the distortion is due to
scattering from the even nucleons.  The appropriate potential to use 
in calculating these distorted waves is unclear.  The KMT treatment 
for the elastic scattering from two particles indicates
that the potential should contain a factor of 1/2.  Using only the
wave functions from this calculation, however, no correction to the
amplitude is ever made. Thus it is not clear if a factor of 1/2, 
unity, or an intermediate value should be used. 
The use of 1/2 gave favorable results when calculating 
the odd-nucleon interaction cross sections in Ref. \cite{gg} and we 
use the same value here.

In calculating the DSF amplitude
the distortion is taken to be from the odd nucleon and one of the even 
nucleons. We use a factor of unity here. This is consistent with the study
of single charge exchange  \cite{june} (in which the distortion is also 
due to one strong and one weak interaction), for which the results were 
best when the full optical potential was used.

\section{Double Spin Flip Calculation}\label{sec:calc}

Let {$\bfk$} and {$\bfkprime$} be the pion's initial and final 
center of mass momenta, respectively,
and {$\left|\bfk\right|=\left|\bfkprime\right|=k$}, 
{$\bfkhat\cdot\bfkprimehat=\cos\theta\,$}.
 Let the two even nucleons have coordinates {$\bfrone$} and
{$\bfrtwo\,$}, spin operators {$\bfsigmaone$} and
{$\bfsigmatwo\,$}, and
wave function {$\chi(\bfrone,\bfrtwo)\,$}. 
The pion's initial and final distorted wave functions are then $\Psip$ and
$\Psim$. Assuming closure over 
the intermediate states and plane wave propagation of the pion
in the intermediate state, the double scattering amplitude is 
given by
$$
\fdsf=
\frac{1}{2\pi^2} \int d\bfrone \, d\bfrtwo \, d\bfq \, 
\chi^{\ast}(\bfrone,\bfrtwo) \, \left[
f_{2}(\bfq,\bfktwo) \, \Psim  \right]  
\, e^{i\bfq\cdot\bfrtwo} \, P(q) \; \times
$$
\eq 
e^{-i\bfq\cdot\bfrone}\,\left[  f_{1}(\bfkone,\bfq)  \, 
\Psip \right]  \, \chi(\bfrone,\bfrtwo) \; ,
\qe
where the pion propagator is
\eq
P(q) = \frac{1}{q^{2}-k^{2}} \; ,
\qe
the pion-nucleon spin flip amplitude is
\eq
f_{i}(\bfqone,\bfqtwo) = i \, \lambda \, v(q_{1})\, v(q_{2}) \, 
\bfsigmai \cdot \left( \bfqone\times\bfqtwo \right)  \; ,
\qe
and the off-shell form factor is
\eq
v(q) = \frac{\alpha^{2}+k^{2}}{\alpha^{2}+q^{2}} \; .
\qe
The quantity  
{$\alpha$} describes the range of the pion-nucleon interaction, which we
take to be 600 MeV/c, 
and {$\lambda$} is derived from the $\pi$N phase shifts \cite{arndt}. 
Since $f_{i}$ is an operator in nucleon spin space, 
we must compute the expectation value of the operator
\eq
O=\left(\bfsigmaone\cdot\bfkone\times\bfq\right)
\left(\bfsigmatwo\cdot\bfq\times\bfktwo\right) \; .
\qe 
For a pure singlet case \cite{gek} 
\eq
\left\langle O \right\rangle _{S=0} = 
- \left( \bfkone\times\bfq \right) \cdot
  \left( \bfq \times \bfktwo \right)
=-q^{2} \left[ \left( \bfkone\cdot\bfqhat \right) 
\left( \bfktwo\cdot\bfqhat \right)-\bfkone\cdot\bfktwo \right] \; . 
\qe
We write
\eq  
\left| \chi(\bfrone,\bfrtwo) \right|^{2} = \rhoone \, \rhotwo
\qe
for the target in its ground state, so that the DSF scattering amplitude
is now
$$
\fdsf =
\frac {\lambda^{2}} {2\pi^{2}} \int d\bfrone \, d\bfrtwo \, d\bfq
\, q^{2} \, v^{2}(q) \, P(q) \, 
e^{i\bfq\cdot\bfrtwo} \, \rhotwo \,
e^{-i\bfq\cdot\bfrone} \, \rhoone \; \times
$$
\eq
v(k_1) \, v(k_2) 
\left[ \left( \bfkone\cdot\bfqhat \right)
\left( \bfktwo\cdot\bfqhat \right)-\bfkone\cdot\bfktwo \right] 
\Psim \, \Psip \; .
\label{eq:fdsf} \qe

As the momenta are operators on the distorted waves, we set
\eq
\bfk_j \equiv -i \, \bfdelj \; ,
\qe
and define
\eq
\psione \equiv  v(k_1)  \Psip \, ; \ \ 
\psitwo \equiv  v(k_2)  \Psim \; .
\qe
We use a partial wave expansion for the pion wave function
\eq
\Psip=
4 \pi \sum_{lm} i^{l} \, Y_{lm}(\bfkhat) \,
Y_{lm}^{\ast}(\bfronehat) \, \Psi_l(kr_1) \; ,
\qe
so that, by use of a double Fourier transform and 
orthonormality of the spherical harmonics,
\eqan
\psione & = &
    \frac {1} {\left( 2\pi \right)^{3}}
    \frac{\alpha^2+k^2} {\alpha^2 - \bfdelone^2} 
    \int d\bfkap \, e^{-i\bfkap\cdot\bfrone} \, 
    \Psi^{(+)}(\bfk,\bfkap)
\\
& = &
    \frac {2}{\pi} \lp\alpha^{2}+k^{2}\rp 
    4 \pi \sum_{lm} i^{l} \, 
    Y_{lm}(\bfkhat) \,
    Y_{lm}^{\ast}(\bfronehat)
    \int dr \, d\kappa  
    \frac {r^{2} \kappa^{2} j_{l}(\kappa r_{1}) j_{l}(\kappa r) 
            \Psil} {\alpha^{2}+\kappa^{2} } \; ,
\qean
where we identify
\eqan
\psi_{l}(kr_{i}) &=&
      \frac{2}{\pi} \lp\alpha^{2}+k^{2}\rp \int dr \, d\kappa 
      \frac {r^{2} \, \kappa^{2} \, j_{l}(\kappa r_{i}) \, 
             j_{l}(\kappa r) \, \Psil} {\alpha^{2}+\kappa^{2}}  \\
& = & \int dr \, r^{2} \, G_{l}(r,r_{i}) \, \Psil \; ,
\qean
and where contour integration gives
\eqan
G_{l}(r,r\p) & \equiv &
    \frac {2}{\pi} \lp\alpha^2+k^2\rp \int_{0}^{\infty} d\kappa 
    \frac {\kappa^{2} \, j_{l}(\kappa r\p) \, j_{l}(\kappa r)}
    {\alpha^{2}+\kappa^{2}} \\
& = & {-\alpha}\lp\alpha^{2}+k^{2}\rp  
    h_{l}^{(+)}(i \alpha r\great) \, j_{l}(i\alpha r\less)
\qean
for {$r\great$} the greater of {$r$} and {$r\p$}. 
In a similar fashion
\eq
\psitwo 
=4 \pi \sum_{lm} \lp-i\rp^{l} Y_{lm}(\bfkprimehat) \,
Y_{lm}^{\ast}(\bfrtwohat) \, \psi_{l}(kr_{2}) \; ,
\qe 
as {$\Psi^{(-) \ast}(\bfk,\bfr)=\Psi^{(+)}(-\bfk,\bfr)\,$}.

Thus Eq.~\ref{eq:fdsf} in the DWIA may be written
$$
\fdsf =
\frac {\lambda^{2}} {2\pi^{2}} \int d\bfrone \, d\bfrtwo \, d\bfq
\, q^{2} \, v^{2}(q) \, P(q) \,
e^{i\bfq\cdot\bfrtwo} \, \rhotwo \,
e^{-i\bfq\cdot\bfrone} \, \rhoone
\; \times
$$
\eq
\left[ \bfdelone\cdot\bfdeltwo - \left( \bfqhat\cdot\bfdelone \right) 
\left( \bfqhat \cdot \bfdeltwo \right) \right]
\psione \, \psitwo \; .
\label{eq:fddsf} 
\qe

The DSF amplitude is twice the value given by Eq. \ref{eq:fddsf}, 
as the process may proceed in two time orders.

\subsection{Evaluation: First Technique}\label{firstway}
While Eq.~\ref{eq:fddsf} may be calculated directly, it is more 
easily evaluated by integration by parts on {$\bfrone$} and 
{$\bfrtwo\,$}.
We find
\eqa
\fdsf & = &
    -\frac {\lambda^{2}} {2\pi^{2}} \int d\bfrone \, d\bfrtwo 
    \, d\bfq \, q^{2} \, v^{2}(q) \, P(q)  \, \psione \, 
    e^{-i\bfq\cdot\bfrone} \, \psitwo \, e^{i\bfq\cdot\bfrtwo}   
\; \times 
\qea
\eq
   \rhopone \, \rhoptwo 
   \left[  \bfronehat\cdot\bfrtwohat 
     - \left( \bfqhat\cdot\bfronehat\right) 
   \left( \bfqhat\cdot\bfrtwohat \right) \right].
\label{eq:bp}\qe
Let us consider each term in brackets separately, and label them
{$F_1$} and {$F_2$} respectively.

The first term is calculated by expanding the wave functions,
exponentials, and $\bfronehat\cdot\bfrtwohat$ in partial waves.  We find
\eq
\fdsfone=
-\frac {\lambda^{2}} {2\pi^{2}} \left( 4\pi \right)^{3} 
\sum_{L l}  \lp 2L+1 \rp
\left(  C_{\,1 \,L \,l}^{0 \,0 \,0} \right)^{2} 
P_{L}(\cos\theta) \, I_{L,L,l,l}(r_{1},r_{2}) \; ,
\label{eq:fdsfone}\qe
where
\eq
I_{L_1,L_2,l_1,l_2}(r_{1},r_{2}) \equiv
\int dr_{1} \, dr_{2} \, r_{1}^{2} \, r_{2}^{2} \, \rhopone \,
\rhoptwo \, \psi_{L_1}(kr_1) \, \psi_{L_2}(kr_2) \,
\bigeta_{l_1,l_2}(r_{1},r_{2}) \; ,
\qe
and
\eqa
\bigeta_{l_1,l_2}(r_{1},r_{2}) & \equiv & 
     \int_{0}^{\infty} dq \, q^{4} \, v^2(q) \, 
     P(q) \, j_{l_1}(qr_{1}) \, j_{l_2}(qr_{2})
\\
& = &
    \frac{i\pi}{2}  k^3 \, j_{l\less}(kr\less) \,
    h_{l\great}^{(+)}(kr\great)
    +\frac{\pi}{4}  \alpha  \lp \alpha^{2}+3k^{2} \rp  
    j_{l\less}(i \alpha r\less) \, 
    h_{l\great}^{(+)}(i \alpha r\great)
\qea
\eq
+\frac {i\pi}{4} \alpha^{2} \left( \alpha^{2} + k^{2} \right) 
\left[ r\less \, j_{l\less}\p(i \alpha r\less) \,
       h_{l\great}^{(+)}(i \alpha r\great) 
   + r\great \, j_{l\less}(i \alpha r\less) \, 
       h_{l\great}^{(+)\prime}(i \alpha r\great) \right]
\label{eta} \qe 
provided {$l\less < (l\great+3)$} 
where {$r\great$} is the
greater of {$r_{1}$} and {$r_{2}$} and {$l\great$} is
{$r\great$}'s respective index.
To calculate the second term of Eq. \ref{eq:bp}, we write
\eq 
\bfqhat\cdot\bfrhat \, e^{i\bfqhat\cdot\bfrhat}
=4 \pi \sum_{lLM} i^{l} \, j_{l}(qr) \,
\left(  C_{1 \,L \,l}^{0 \,0 \,0} \right) ^{2}
Y_{LM}(\bfqhat) \, Y_{LM}^{\ast}(\bfrhat).
\label{eq:qdotreqdotr} 
\qe
Expanding the functions as before, we find
\eq
\fdsftwo =
\frac{\lambda^{2}}{2\pi^{2}} \left( 4\pi \right)^{3}
\sum_{l_{1} l_2 L} i^{l_{2}-l_{1}} \lp 2L+1\rp
\left( C_{1 \, L \, l_{1}}^{0 \, 0 \, 0} \right) ^{2}
\left( C_{1 \, L \, l_{2}}^{0 \, 0 \, 0} \right) ^{2}
P_{L}(\cos\theta) \, I_{L,L,l_{1},l_{2}}(r_{1},r_{2})   
\; .\qe
Thus we have
$$
\fdsf =
    32 \pi \lambda^{2} \sum_{L} \lp 2L+1 \rp  P_{L}(\cos\theta)
\; \times
$$
\eq
    \lc
    \sum_{l_{1} \neq l_{2}} i^{l_{2}-l_{1}}
    \left( C_{1 \, L \, l_{1}}^{0 \, 0 \, 0} \right)^{2}
    \left( C_{1 \, L \, l_{2}}^{0 \, 0 \, 0} \right)^{2}
    I_{L,L,l_{1},l_{2}}(r_{1},r_{2})
    + \sum_{l} \left[
      \left( C_{1 \, L \, l}^{0 \, 0 \, 0} \right)^{4}
    - \left( C_{1 \, L \, l}^{0 \, 0 \, 0} \right) ^{2} \right]
      I_{L,L,l,l}(r_{1},r_{2})
    \rc \; .
\qe
We may further reduce this expression with Racah algebra,
$$
\fdsf  =  
    32 \pi \lambda^{2} \sum_{L} \lp 2L+1 \rp
    P_{L}(\cos\theta)
\; \times
$$
$$
    \lcurly -2 \left(  C_{1 \;L \;L-1}^{0 \;0 \;\;\;\;0} \right)^{2}
    \left( C_{1 \;L \;L+1}^{0 \;0 \;\;\;\;0} \right)^{2}
    I_{L,L,L-1,L+1}(r_{1},r_{2})
\quad\quad\quad\;
$$
$$
    + \left[  \left(  C_{1 \;L \;L-1}^{0 \;0 \;\;\;\;0} \right)^{4}
    - \left( C_{1 \;L \;L-1}^{0 \;0 \;\;\;\;0} \right)^{2} \right]
    I_{L,L,L-1,L-1}(r_{1},r_{2})
$$
\eq
\quad\:\:
    +\left[ \left(  C_{1 \;L \;L+1}^{0 \;0 \;\;\;\;0} \right)^{4}
    -\left( C_{1 \; L \;L+1}^{0 \;0 \;\;\;\;0}\right) ^{2} \right]
    I_{L,L,L+1,L+1}(r_{1},r_{2})
    \;\rcurly  \; .
\qe
As the first two indices on the integral term, $I$, are
the indices for the pion's wave function, the angular momentum of the pion
is conserved as it must be for the elastic interaction with a spin-zero
system.

\subsection{Evaluation: Second Technique}\label{secondway}

Alternately, one may directly calculate Eq. \ref{eq:fddsf}. This
calculation is straightforward, but lengthy. 
We use the covariant components of the operator $\bfdel$ in the
spherical basis.  For {$\sigma = \lp -1,0,1 \rp \, $},	
\eq
\bfdelone \cdot \bfdeltwo =
 \sum_{\sigma} \lp -1 \rp^{\sigma} \, 
 \nabla_{\sigma}^{(1)} \, \nabla_{-\sigma}^{(2)}
\; ,
\label{eq:aboveone}
\qe
$$
\left( \bfqhat\cdot\bfdelone \right) 
\left( \bfqhat\cdot\bfdeltwo \right)
=
     \sqrt{4\pi} \sum_{\sigma \sigma\p}
     \left[  C_{1 \:1 \:0}^{0 \,0 \,0} \,
     C_{\,1 \:1 \;\;0}^{\sigma \, \sigma\p \,0} \, Y_{00}(\bfqhat)
     +\frac{1}{\sqrt{5}} \sum_{M}(-1)^M \, C_{1 \:1 \:2}^{0 \,0 \,0} \,
     C_{\,1 \:1 \;\;\;2}^{\sigma \, \sigma\p \, M} \, Y_{2M}(\bfqhat) 
     \right]  \nabla_{-\sigma}^{(1)} \, \nabla_{-\sigma\p}^{(2)}
$$
\eq
=
  \frac13 \bfdelone\cdot\bfdeltwo
     + \sqrt{\frac{8\pi}{15}} \sum_{\sigma \sigma\p M}(-1)^M \, 
     C_{\,1 \:1 \;\;\;2}^{\sigma \, \sigma\p \, M} \, Y_{2M}(\bfqhat)
     \, \nabla_{-\sigma}^{(1)} \, \nabla_{-\sigma\p}^{(2)}
   \; .
\label{eq:abovetwo} 
\qe
Let $\fdsfs$ be the contribution to $\fdsf$ from Eq.~\ref{eq:aboveone}
and let $\fdsfd$ be the contribution from the second
term in Eq.~\ref{eq:abovetwo}, so that
\eq
F(\bfk,\bfkprime)=
\frac{2}{3}F_{S}(\bfk,\bfkprime)+F_{D}(\bfk,\bfkprime) \; .
\qe
From Reference~\cite{angmom} we have
\eqa
\nabla_{\sigma} \, Y_{\lambda\mu}(\bfrihat) \, \psi_{\lambda}(kr_{i})  
& = &
      \sqrt { \frac {\lambda+1} {2 \lambda+3} }  
      C_{\lambda \,\,1 \:\lambda+1}^{\mu \,\sigma \,\mu+\sigma} \,
      F_{\lambda}(r_{i}) \, Y_{\lambda+1 \, \mu+\sigma}(\bfrihat)
\\
& & \mbox{}
      -\sqrt { \frac {\lambda} {2\lambda-1} }
      C_{\lambda \,\,1 \;\lambda-1}^{\mu \,\sigma \,\mu+\sigma} \,
      G_{\lambda}(r_{i}) \, Y_{\lambda-1 \, \mu+\sigma}(\bfrihat) \; ,
\qea
\eq
F_{\lambda}(r_{i})  \equiv  
     \frac {d\psi_{\lambda}(kr_i)} {dr_i}
     -\frac {\lambda} {r_i}  \psi_{\lambda}(kr_{i}) \, ; \ \ 
G_{\lambda}(r_{i})  \equiv 
     \frac {d\psi_{\lambda}(kr_i)} {dr_i}
     +\frac {\lambda+1} {r_i}  \psi_{\lambda}(kr_{i}) \; .
\qe
If we proceed in a manner similar to that in Section~\ref{firstway}, we
find
$$
F_{S}(\bfk,\bfkprime)=
32 \pi \lambda_{f}^{2}
\sum_{l} P_{l}(\cos\theta) \int dr_{1} \, dr_{2} \, r_{1}^{2} \, 
r_{2}^{2} \, \rhoonesq \, \rhotwosq \; \times
$$
\eq
\left[ \lp l+1 \rp \bigeta_{l+1,l+1}(r_{1},r_{2}) \, 
F_{l}(r_{1}) \, F_{l}(r_{2}) 
+ l \, \bigeta_{l-1,l-1}(r_{1},r_{2}) \, G_{l}(r_{1}) \, 
G_{l}(r_{2}) \right]
\qe
and
$$
F_{D}(\bfk,\bfkprime) = 
 -32 \pi \sqrt{\frac{2}{3}} \lambda_{f}^{2}
 \sum_{l} \lp -1 \rp^{l}
 P_{l}(\cos\theta) \int dr_{1} \, dr_{2} \, r_{1}^{2} \, r_{2}^{2} 
 \, \rhoonesq \, \rhotwosq \; \times
$$
$$
 \lbrac
 \lp 2l+3 \rp  \lp l+1 \rp  \bigeta_{l+1,l+1}(r_{1},r_{2})
 \, F_{l}(r_{1}) \, F_{l}(r_{2}) \, 
 C_{l+1 \: l+1 \: 2}^{\,\,0 \,\;\;\;\;0 \:\:\:\:0} \, 
 {\tiny
 \left\{
 \begin{array}
 [c]{ccc}%
 l & l+1 & 1\\
 2 & 1 & l+1
 \end{array}
 \right\} }
$$
$$
 +l  \lp 2l-1 \rp  \bigeta_{l-1,l-1}(r_{1},r_{2}) \,
 G_{l}(r_{1}) \, G_{l}(r_{2}) \, 
 C_{l-1 \: l-1 \: 2}^{\,\,0 \,\;\;\;\;0 \:\:\:\:0} \, 
 {\tiny
 \left\{
 \begin{array}
 [c]{ccc}%
 l & l-1 & 1\\
 2 & 1 & l-1
 \end{array}
 \right\} }
$$
\eq
 +2\sqrt{l  \lp l+1 \rp  \lp 2l-1 \rp  \lp 2l+3 \rp} 
 \bigeta_{l-1,l+1}(r_{1},r_{2}) \,
 G_{l}(r_{1}) \, F_{l}(r_{2}) \, 
 C_{l-1 \: l+1 \: 2}^{\,\,0 \,\;\;\;\;0 \:\:\:\:0} \, 
 {\tiny
 \left\{
 \begin{array}
 [c]{ccc}%
 l & l-1 & 1\\
 2 & 1 & l+1
 \end{array}
 \right\} }
 \rbrac
\qe
where $\bigeta_{l_1,l_2}(r_{1},r_{2})$ is as previously
defined.

\section{Results}\label{sec:results}

We performed calculations with the density of {\heth} taken from 
the solution to the Faddeev Eqs. \cite{friar}.  
The two treatments presented in the previous section were evaluated
and agree to within the expected numerical accuracy. The first method
depends on the numerical calculation of the derivatives of the
three-body density while the second does not.  The derivative 
operators act on the pion wave function in the second method while 
the first method has a more direct form to calculate, i.e., it is 
expressed as an expectation value of a {\em local} operator.
It is easier to see how to include the $\delta$-function
effect, discussed shortly, with the second method.

Figure~\ref{gr:ddsf} shows the results of adding the DSF to the
basic optical model for all four scattering cases.  
The dashed lines correspond to the optical model only and the solid 
lines include the DSF term.  We see that the addition of the DSF 
term is indeed significant at large angles for the even nucleon cases. 
However, it does not give the structure seen in experiment.

\begin{figure}
\epsfysize=6.0in 
\epsffile{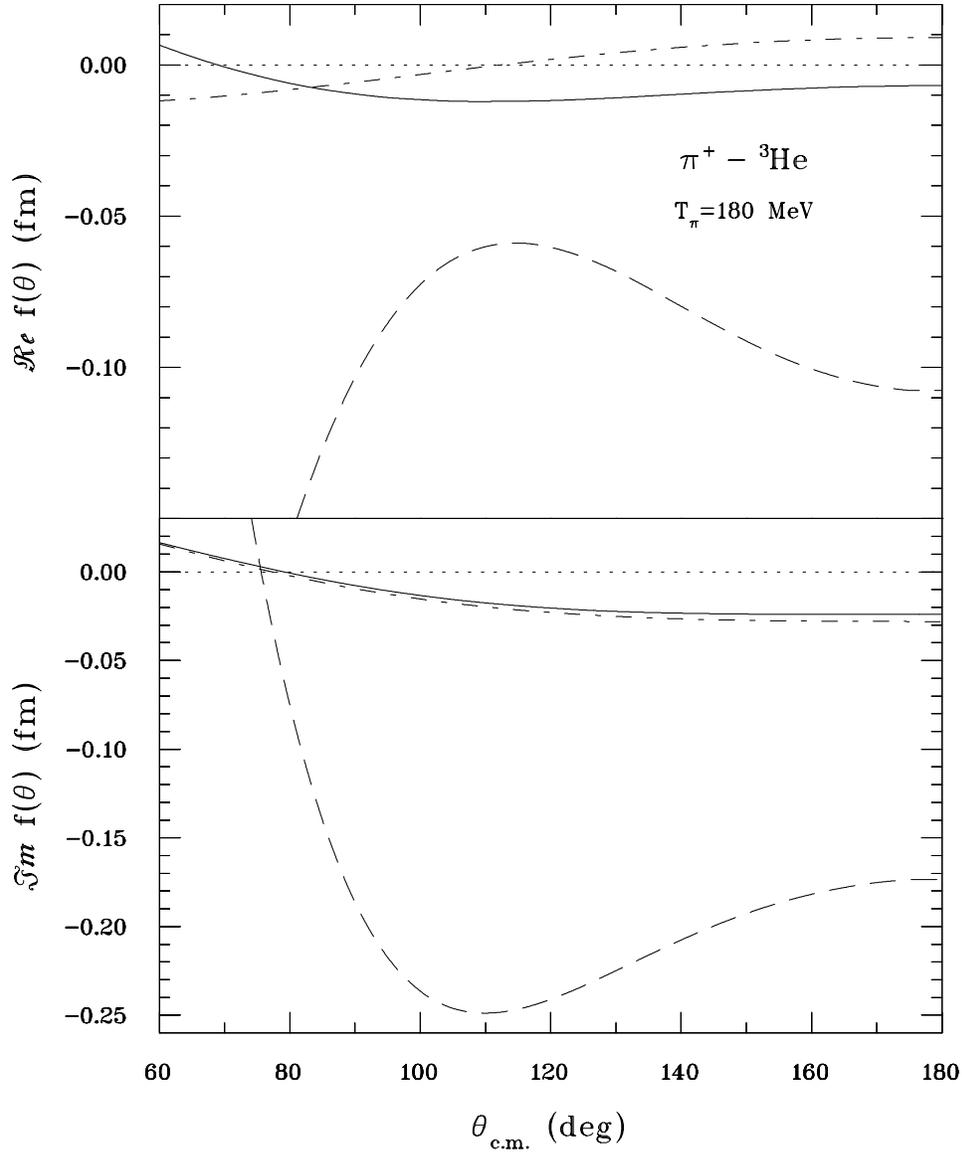}
\caption{Scattering amplitudes for {\piphe}.
The dashed line represents $f(\theta)$ from the optical model,
the solid line the double spin-flip
mechanism (solid line),  and the dash-dotted line the DSF with
the $\delta$-function removed.}
\label{gr:amp}\end{figure}

Figure~\ref{gr:amp} compares the scattering amplitude of the DSF to 
that of the basic optical model for {\piphe}.  The dashed line 
represents $f(\theta)$ from the optical model only, the solid line 
the DSF amplitude, and the dash-dotted-line the DSF amplitude with the
$\delta$-function removed (discussed below).  We see
that in the region of the first minimum both the real and imaginary parts 
of the DSF amplitude are passing through zero, and thus have little effect 
in this region. 

One may speculate that a cancellation between the two mechanisms
might produce a minimum at 130{\degree} and that the present calculation
does not have the correct phase.  To check this possibility, we
introduced an arbitrary phase difference between the two terms, 
but still found no case which gave the characteristic minima.  

\begin{figure}
\epsfysize=6.0in 
\epsffile{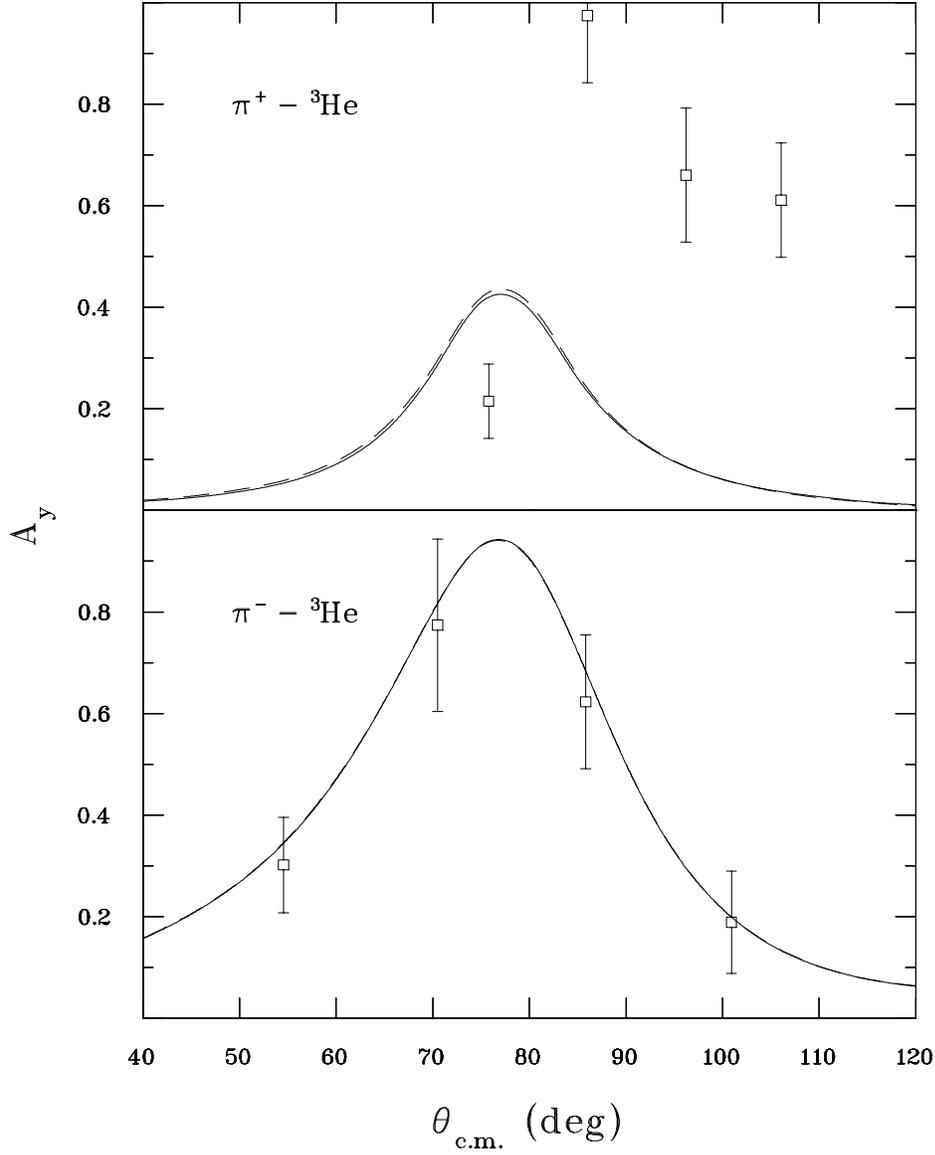}
\caption{Asymmetries for $\pi^+\ ^3$He and  $\pi^-\ ^3$He scattering from 
the optical model only (dashed line) and with the DSF term (solid line).
The data are from Refs. \protect\cite{piplus} and 
\protect\cite{piminus}. For {\pip} there are additional data
points at forward angles which are negative (not shown).}
\label{gr:asypm}\end{figure}

The asymmetries are shown in Fig.~\ref{gr:asypm}. Since polarization
phenomena are typically sensitive to small corrections, one might
expect important corrections from the DSF; but we note that its 
addition has little effect since the asymmetry is only large around
the minimum in the cross section where the DSF amplitude vanishes.

It has long been known that there is a $\delta$-function
in the s-wave (nucleon-nucleon) portion of the one-pion-exchange
potential (see e.g. Ref. \cite{gl} for a discussion of this 
$\delta$-function and its removal).  This $\delta$-function also
exists in the p-wave--p-wave part of pion double scattering
\cite{ee,bb}, especially
visible in double charge exchange \cite{gek,dcx}.

If we refer back to Eq.~\ref{eq:fdsf}, we see that the $\delta$-function
piece of the DSF amplitude results from the monopole portion of the
expression 
$\left[ \left( \bfkone\cdot\bfqhat \right)
\left( \bfktwo\cdot\bfqhat \right)-\bfkone\cdot\bfktwo \right]\,$.  In
Section~\ref{secondway} we explicitly expanded this expression in terms
of the $\bfdel$ operator.  Thus from the second treatment it is easy
to see that the $\delta$-function should be removed from the $\fdsfs$ term 
only. To do so, we make the replacement
\eq
\frac {q^{2}} {q^{2}-k^{2}} \rightarrow 
\left[ \frac {q^{2}} {q^{2}-k^{2}} -1 \right] 
=\frac{k^2}{q^2-k^2} \; ,
\qe
which is equivalent to replacing $q^4$ with $k^2q^2$ in
the integral expression for $\bigeta_{l_1,l_2}(r_1,r_2)$ in
Eq.~\ref{eta}.
The term $\fdsfd$ remains unchanged.

As seen in Fig.~\ref{gr:amp}, this correction essentially reverses the
sign of the real part of the scattering amplitude, while having a small
effect on the imaginary part.
However, it is the imaginary portion of the scattering amplitude that
dominates
the cross section at large angles where the DSF is of significance.
Thus the $\delta$-function correction has a minimal effect on the
scattering cross section, decreasing it by less than 4 percent.
In principle this correction should also be made in the basic
optical model, akin to the Lorentz-Lorenz effect \cite{ee,ehm}
at resonance energies. However,
such a consideration is beyond the scope of this paper.

\begin{figure}
\epsfysize=6.0in 
\epsffile{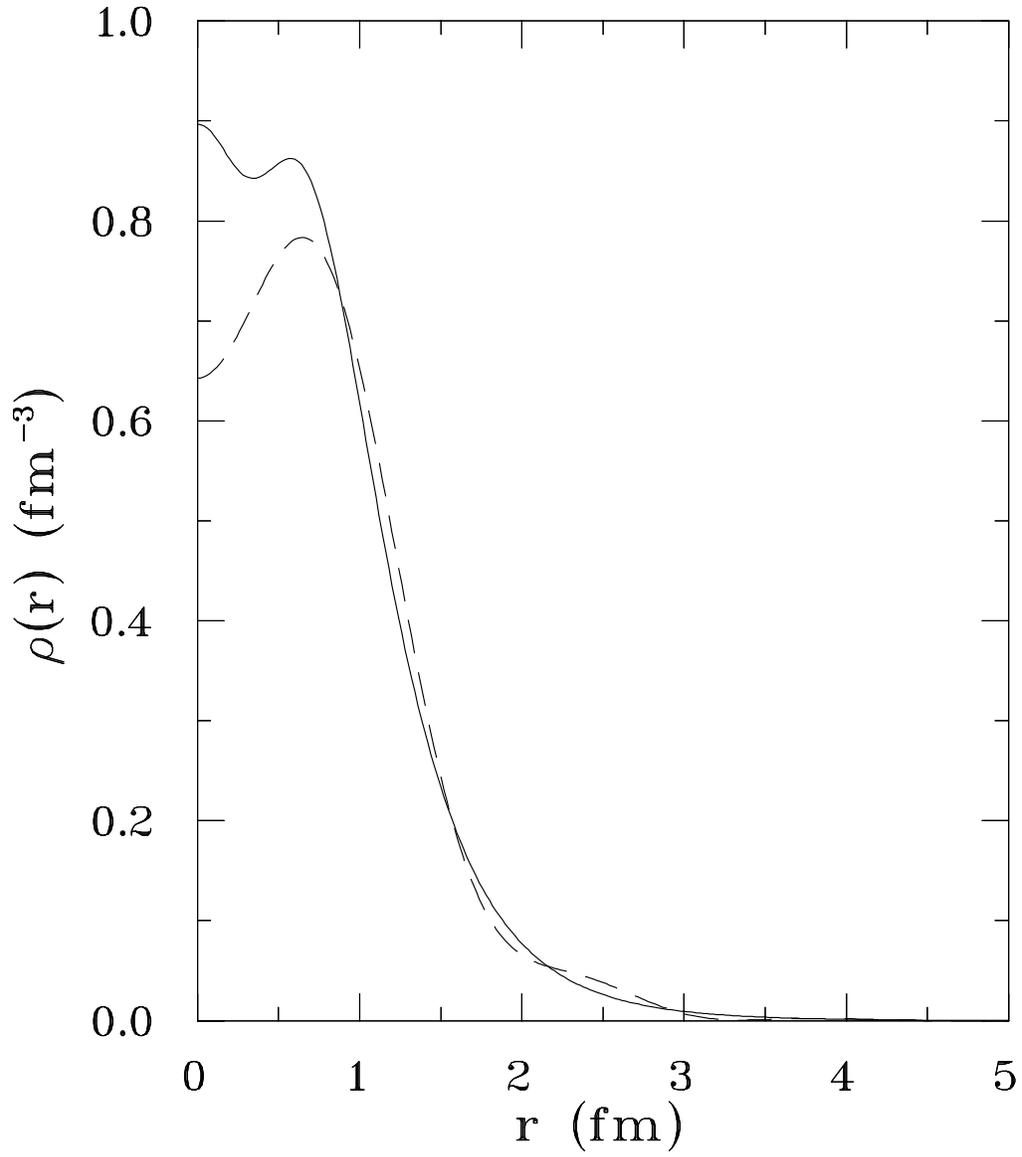}
\caption{Comparison of the Faddeev (solid line) and electron
scattering densities (dashed line).}
\label{gr:rhos}\end{figure}

It is known that the Faddeev densities, while representing the  
exact solution to the three-nucleon system expressed in terms of
nucleon degrees of freedom, do not provide a completely accurate
description of the electron-scattering cross sections.  The
problem is clearer at high momentum transfer near the first
zero of the angular distribution \cite{electronscatt}.  The 
difference is often ascribed to meson exchange currents \cite{mec}.
However, in order to explore the sensitivity of the results
to the density used, we also performed calculations using the
charge densities of Ref. \cite{mccarthy} corrected for the finite 
size of the proton using the proton charge parameterization from 
Ref. \cite{hoehler}.  The two densities are compared in Fig.  
\ref{gr:rhos}.  Only the proton density was changed, the neutron
density remaining that of the Faddeev calculations.

\begin{figure}
\epsfysize=6.0in 
\epsffile{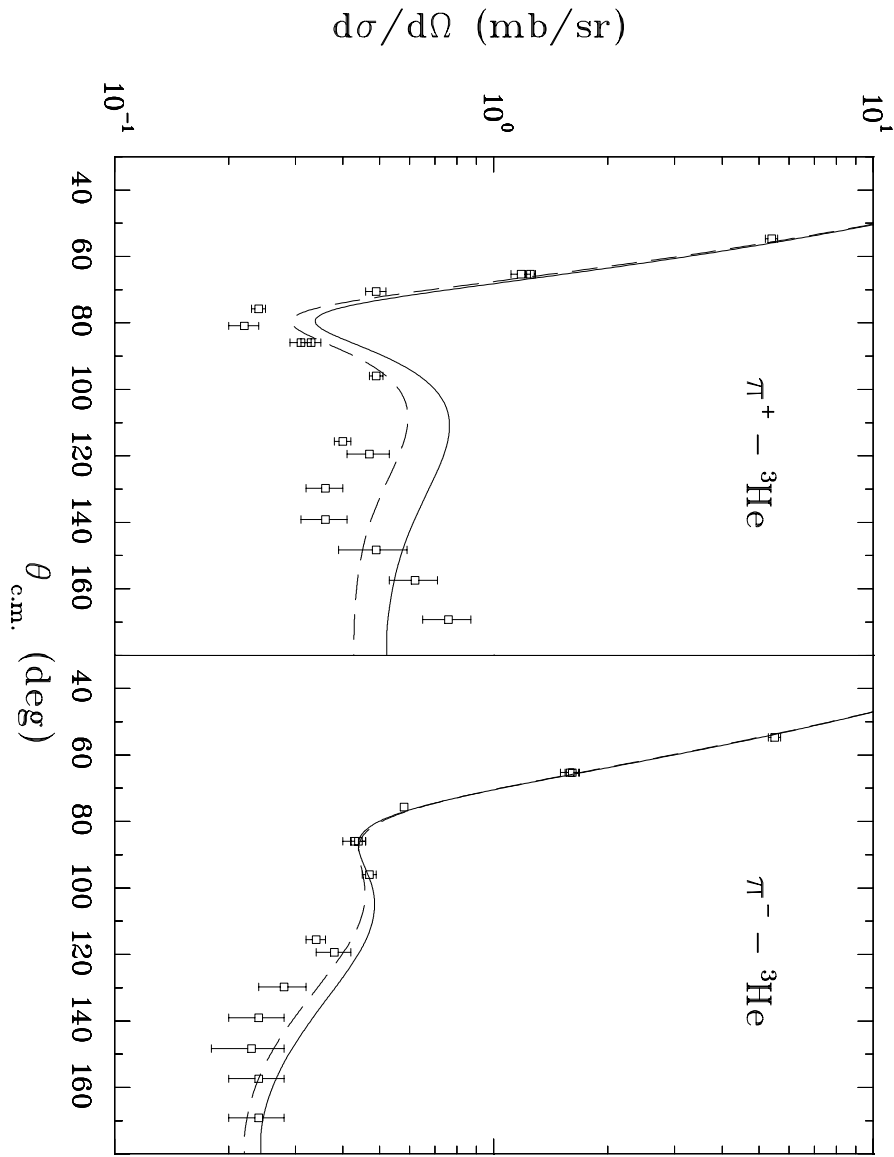}
\caption{Comparison of the results of the optical model with the
Faddeev density (solid line) and the electron scattering density
 (dashed line). The radius for the electron
scattering density has been
rescaled the same r.m.s. value as the Faddeev
density.  If no such rescaling is done, the difference is
considerably less.}
\label{gr:comrho}\end{figure}

Figure \ref{gr:comrho} shows the comparison of the results of 
\piphe\ scattering for the case in which the electron-scattering  
density was adjusted to have the same r.m.s. radius as the 
Faddeev density.  
The introduction of the electron scattering density at most shifts the 
cross section downward, and has little effect on the shape of the 
cross section at large scattering angles.  
For the calculation in which the two densities 
were used in their unmodified form the difference is even less. 
Thus it seems unlikely that the minimum is a form-factor effect, 
at least within the span of currently accepted functions.  One 
can find a fit to the data by allowing an arbitrary form factor 
variation \cite{zhao} but the density which results appears to
be unphysical, having essentially no support for small values of
the radial variable, i.e., it has a complete hole in the center.

\section{Conclusions}\label{sec:concl}
We find that the double-spin-flip correction gives a non-negligible
contribution to the large angle scattering cross section 
in the case of \piphe.
It should be included in any serious attempt to explain quantitatively
the back-angle data on the three-nucleon system, and perhaps on
other nuclei as well.
However, because of the shape and phase of this coherent contribution,
it does not appear that it can explain the dip and subsequent rise
seen in the case of two strong scatters.

As one possible direction for further work to explain the effect,
we observe that if the KMT factor is set to unity, while the 
agreement of the forward cross section with data is completely ruined,
there is a slight dip at the appropriate back
angle. This suggests than an improvement to the basic optical model may
lead to the resolution of this issue. An approach which 
may offer hope is that of Garcilazo \cite{garcilazo}, but such a
treatment involves a complete reformulation of the scattering theory.

The single energy shift used in the present calculations 
might also cause problems.  In Ref. 
\cite{kg} it was found that the shift in energy was dependent
on the pion-nucleus angular momentum which is an alternate
expression of the results of earlier work in which it was a
function of momentum transfer \cite{schmit}.  

These two observations lead one to think that the next
appropriate step might be a re-examination of the scattering theory
to provide a more detailed representation of the physics.

One of us (SLC) wishes to thank the University of Maryland Physics
Department, especially the Theoretical Quarks, Hadrons, and Nuclei
Group, for their hospitality. We wish to thank W. B. Kaufmann
for very useful comments on the comparison of the two formalisms
presented. This work was supported by the U. S. Department of Energy.

\newpage

\begin{thebibliography}{99}

\bibitem{scott} S. K. Matthews \ea, Phys. Rev. C {\bf 51}, 2534 (1995); 
W. J. Briscoe \ea, Nucl. Phys. {\bf A553}, 585c (1993).

\bibitem{ktb} S. S. Kamalov, L. Tiator, and C. Bennhold, 
Phys. Rev. C {\bf 47}, 941 (1993). 

\bibitem{gg} W. R. Gibbs and B. F. Gibson, Phys. Rev. C {\bf 43}, 
1012 (1991).

\bibitem{data} C. Pillai, D. B. Barlow, B. L. Berman, W. J. Briscoe, A.
Mokhtari, B. M. K. Nefkens, M. E. Sadler, Phys. Rev. C {\bf 43}, 1838
(1991); B. M. K. Nefkens, W. J. Briscoe, A. D. Eichon, D. H. Fitzgerald,
A. Mokhtari, J. A. Wightman, M. E. Sadler, Phys. Rev. C {\bf 41}, 2770
(1990). 

\bibitem{burleson} K. S. Dhuga \ea, Phys. Rev. C {\bf 35}, 
1148 (1987).

\bibitem{schiff} L. I. Schiff, Phys. Rev. {\bf 133}, B802 (1964).

\bibitem{scotttogibbs} S. K. Matthews, private communication to W. R.
Gibbs.

\bibitem{franco} V. Franco, Phys. Rev. {\bf 9}, 1690 (1974).


\bibitem{kg} W. B. Kaufmann and W. R. Gibbs,  Phys. Rev. C {\bf 28}, 
1286 (1983).

\bibitem {gak} W. R. Gibbs, Li Ai and W. B. Kaufmann, Phys. Rev.
C {\bf 57}, 784 (1998).

\bibitem{elastic} M. Nuseirat, M. A. K. Lodhi and 
W. R. Gibbs, Phys. Rev. C {\bf 58}, 314 (1998).

\bibitem{garg} H. Garcilazo and W. R. Gibbs, Nuc. Phys. {\bf A381},
487 (1982).

\bibitem{oset}  E. Oset, M Khankhasayev, J. Nieves, H. Sarafian,
and M. J. Vicente-Vacas, Phys. Rev. C {\bf 46}, 2406 (1992).

\bibitem{dcx} M. Nuseirat, M. O. El-Ghossain, M. A. K. Lodhi, 
W. R. Gibbs and W. B. Kaufmann, Phys. Rev. C {\bf 58}, 2292 (1998).

\bibitem{kmt} A. K. Kerman, H. McManus and R. M. Thaler, Ann. Phys. 
{\bf 8}, 551 (1959).

\bibitem{june} M. L. Dowell, W. Fong, J. L. Matthews, H. Park, M. 
Wang, M. E. Yuly, E. R. Kinney, P. A. M. Gram, D. A. Roberts, 
G. A. Rebka, Jr., Phys. Rev. C {\bf 51}, 1551 (1995).  

\bibitem{arndt} R. A. Arndt and R. Roper, SAID interactive 
dial-in program (unpublished).

\bibitem{gghsk}  W. R. Gibbs, B. F. Gibson, A. T. Hess, G. J. Stephenson 
and W. B. Kaufmann, Phys. Rev. C {\bf 13}, 2433 (1976); 
R. Mach, Nucl. Phys. {\bf A205}, 56 (1973); 
R. Landau, Phys. Rev. C {\bf 15}, 2127 (1977); 
R. Landau, Ann. Phys. {\bf 92}, 205 (1975); and Ref. \cite{kg}.

\bibitem{gek} W. R. Gibbs, M. Elghossain and W. B. Kaufmann, 
Phys. Rev. C {\bf 48}, 1546 (1993).

\bibitem{angmom}
D. A. Varshalovich, A. N. Moskalev, V. K. Khersonskii, {\it Quantum 
Theory of Angular Momentum, } World Scientific (1988).

\bibitem{friar}  
J. L. Friar, B. F. Gibson and G. L. Payne, Phys. Rev. C {\bf 35}, 
1502 (1987); J. L. Friar, B. F. Gibson, 
C. R. Chen and G. L. Payne, Phys. Lett. {\bf 161B}, 241 (1985).

\bibitem{piplus} M. A. Espy \ea, {\it Phys. 
Rev. Lett.} {\bf 76}, 3667 (1996).

\bibitem{piminus} M. A. Espy \ea , Phys. Rev. C {\bf 56}, 2607 
(1997).

\bibitem{gl} W. R. Gibbs and B. Loiseau, Phys. Rev. C {\bf 50}, 
2742 (1994).

\bibitem{ee} M. Ericson and T. E. O. Ericson, Ann. Phys. {\bf 36}, 
323 (1966).

\bibitem{bb} G. Baym and G. E. Brown, Nucl. Phys. {\bf A247}, 
395 (1975).

\bibitem{ehm} J. M. Eisenberg, J. H\"ufner and E. J. Moniz,
Phys. Lett. {\bf 47B}, 381 (1973).

\bibitem{electronscatt} A. Amroun, V. Breton, J. M. Cavedon, B. Frois, 
D. Goutte, F. P. Juster, P. Leconte, J. Martino, Y. Mizuno, X. H. Phan,
S. K. Platchkov and I. Sick, Nucl. Phys. {\bf A579}, 596 (1994).

\bibitem{mec} Ch. Hajduk, P. U. Sauer and W. Strueve, Nucl. Phys.
{\bf A405}, 581 (1983); R. Schiavilla, V. R. Pandharipande and
R. B. Wiringa, Nucl. Phys. {\bf A449}, 219 (1986).

\bibitem{mccarthy} J. S. McCarthy, I. Sick, R. R. Whitney, Phys. Rev. C
{\bf 15}, 1396 (1977).

\bibitem{hoehler} G. H\"{o}hler, E. Pietarinen, I. Sabba-Stefanescu, F.
Borkowski, G. G. Simon, V. H. Walther, R. D. Wendling, Nucl. Phys. 
{\bf B114}, 505 (1976).

\bibitem{zhao}  Q. Zhao, Thesis, New Mexico State University (1997)

\bibitem{garcilazo} H. Garcilazo, Phys. Lett. {\bf 82B}, 332 (1979).

\bibitem{schmit} J. Maillet, J. P. Dedonder and C. Schmit,
Nucl. Phys. {\bf A316}, 267 (1979) and references therein.

\end{thebibliography}
\end{document}